# Preparation and characterization of Fe-incorporated TiO$_2$ thin films: A study of optical constants and dispersion energy parameters

Tapash Chandra Paul[1,2], Jiban Podder[2*], Lincoln Paik[3]

[1]Department of Physics, Jagannath University, Dhaka-1100, Bangladesh
[2]Department of Physics, Bangladesh University of Engineering and Technology, Dhaka-1000, Bangladesh
[3]Department of Physics, Bangladesh University of Textiles, Dhaka-1208, Bangladesh

*E-mail: jpodder59@gmail.com, ORCID ID: 0000-0002-4171-4785.

**Abstract**

In this study, we have delineated the preparation and influence of Fe doping on microstructural and optical properties of the pristine and Fe-incorporated TiO$_2$ thin film with different Fe concentrations (0, 2, 4, 6, and 8 at.% ). The samples are prepared by easier and cost-saving way of spray pyrolysis technique (SPT) using Ti(OCH$_2$CH$_2$CH$_2$CH$_3$)$_4$ as a precursor of mother material. The effect of Fe in the microstructure and phase formation of TiO$_2$ thin films is investigated by XRD analysis. XRD investigation depicts that the pristine product corresponds to anatase phase of TiO$_2$ and remains uncontaminated with addition of 2 at.% Fe impurity. It is also observed that Fe introduces a phase transition from anatase to rutile after adulterating more Fe contents (4, 6 and 8 at.%). In order to study optical characteristics, UV–vis spectroscopy has been employed which revels that UV absorption for the Fe incorporated TiO$_2$ products are noticed to move to a longer wavelength (red shift) and Fe contents lessen bandgap energy from 3.81eV (0 at.% Fe) to 3.70 eV (8 at.% Fe) of the TiO$_2$ thin films. The impact of Fe on the optical constants such as refractive index, complex dielectric constants, tanδ, VELF and SELF, dispersion parameters of obtained titanium dioxide samples has been studied. The results display that Fe influences the structural and optical characteristics significantly.

# 1 Introduction

With the development of the applications such as LEDs [1], supercapacitor [2], gas sensor [1], anti-reflective coatings [3], optical waveguides [4], optoelectronics [5], photonic crystals [6] and semiconductor-based device structures [7], transition conducting oxide (TCO) thin filmshave recently grabbed the concentration of the scientists as a prospective contestant. Among the oxide materials, there is a considerable research focus on $TiO_2$ since it becomes one of the fascinating transition metal oxides which has enticing physical and chemical features due to its broad transmittance (~80 %), a wideband gap, high refractive index, fairly good thermal and chemical stabilities [8][9]. The $TiO_2$ crystalline structure has three different phases includinganatase, rutile, and brookite.$TiO_2$ having anatase phase as a TCO has numerous advantages due to its relative low effective mass, low cost and stability in a hydrogen plasma atmosphere which is used to produce solar cells [10]. There is an important evolution to refrain the microstructural and optical properties of $TiO_2$ in order to utilize a highest solar spectrum for the different applications to resolve our environmental problems [11]. Doping with more foreign elements (Zn, Mn, Fe, Ni, Co) of $TiO_2$ nanoparticles affects their optical and structural properties which allows to adapting them for a variety of applications [12]. Kimet al.[13] investigated the formation of $TiO_2$ doping with Fe content by mechanical alloying and represented the photocatalytic characteristics and microstructural analysis. They noticed that Fe shows a remarkable impact on the bandgapenergies which varies 2.5–2.9 eV. Patil et al. [14] used spray pyrolysis route to synthesize the Ni-doped titanium dioxide and reported that Ni doping changed the surface morphology of the deposited products. Houng et al. [10] employed RF magnetron sputtering for the synthesis of $TiO_2$ samples and they revealed that the incorporation of Mo enhances the crystallization and increases the grain size. We have taken Fe as a dopant element because the size of $Fe^{3+}$ (0.64 Å) [15] and $Ti^{4+}$ (0.68 Å)[16] is almost similar and it has a static half-filled $d^5$ electronic configuration [17]. However the presence of Fe as an intrinsic impurity and extrinsic dopant has influence on structural, optical and electrical properties [18].

In view of doping applications, number of techniques have been reportedin the literature for the fabrication of Fe-doped $TiO_2$ thin films which includes pulsed laser deposition method [19], spin coating technique [20], magnetronsputtering [21], electrochemical deposition [22], hydrothermal method [23], sol-gel [24], chemical bath deposition (CBD) [25] etc. In this article we have employed the SPT for numerous advantages. This method is simple and cost effective to synthesize in small dimensions and at a large areas and produced samples are almost uniform and reproducible [26].

The current research focuses on the formation of titanium dioxide thin films as a function of Fe and impact of Fe contents on microstructural and optical characteristics. We have employed W-H and H-W techniques along with Scherrer formula to find the numerous microstructural parameters taking into consideration line broadening. To best of our concern, no systematic reports on the measurement of optical parameters and dispersion energy parameters as function of Fe impurities have been published. Thus the present researchers have given a deep attention on the on the subject of $E_g$, refractive index, dielectric contants, dilectric loss, average values of oscillator strength, VELF and SELF of the samples. The observations reflect that the as-prepared products might have a good potential as an emersing aspirant for optoelectronic applictions.

## 2 Experimental details

### 2.1 Materials

The preliminary solution for the preparation of pristine $TiO_2$ thin film consisted of Titanium butoxide of 0.1M concentrations. The spray pyrolysis system used the ferric chloride as the raw chemical of dopant with the starting solution to obtain the Fe doped products. A little bit of hydrochloric acid (nearly 1.2 mL) was dissolved with both solutions to get consistent solutions. The purity of the materials is 97%, 98%, and 37% respectively. Deionized water was used in all cases of the current research.

### 2.2 Synthesis of $TiO_2$ thin films:

The deposition of pristine and Fe doped samples has been carried out in a simple and low-cost fume chamber as illustrated in **Figure 1** [27] and the detail of the deposition technique is presented in **Table 1**.

**Table 1. Detail of the $TiO_2$ deposition process.**

| Deposition parameters | Optimum value |
| --- | --- |
| Substrate | microscope slides |
| Base pressure | 0.5 bar |
| Deposition time | 10±2 min. |
| Deposition rate | 0.5 mL/min |
| Nozzle to substrate seperation | 30 cm |
| Amount of solution | 100 mL |
| Doping (Fe) concentration (at. %) | 2.0, 4.0, 6.0, 8.0 |
| pH | 5.5 |

In this method, first, a solution of initial precursor material titanium butoxide with appropriate concentration (0.1 M) is obtained into 100 mL de-ionized water in a beaker at room temperature. The initial mixture has been stirred for about one to two hours employing a magnetic stirrer to achieve a harmonized solution. In addition, 1.2 ml amount of HCl was mixed with preceding solution that makes it aqueous. In parallel way, the desired amounts of doping material ferric chloride based on the Fe concentration varied in the range 0–8 at% with an interval 2 is adopted to the resulting solution prior to stirring the solution thoroughly to have the doped products. The glass slides used as substrates during the deposition process are introduced after cleaning employing acetone and de-ionized water successively to eradicate pollutant. The substrate temperature remains 450 ºC in the time of spraying controlled with the help of a Copper- Constantan thermocouple. The synthesis process has been displayed in Figure 2 schematically.

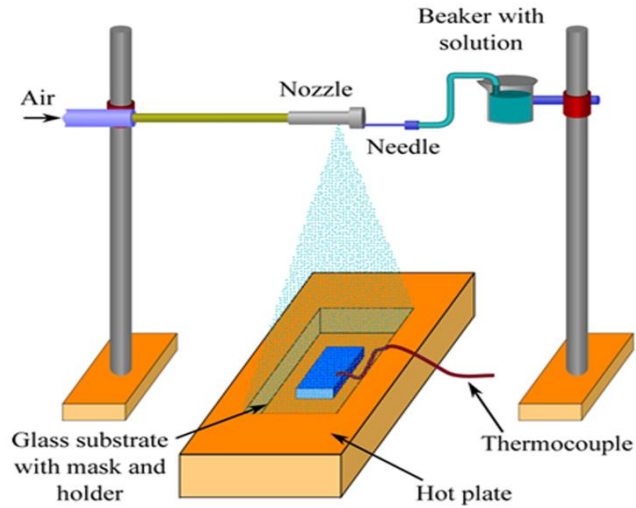

**Figure 1. Experimental setup of the spray pyrolysis technique.**

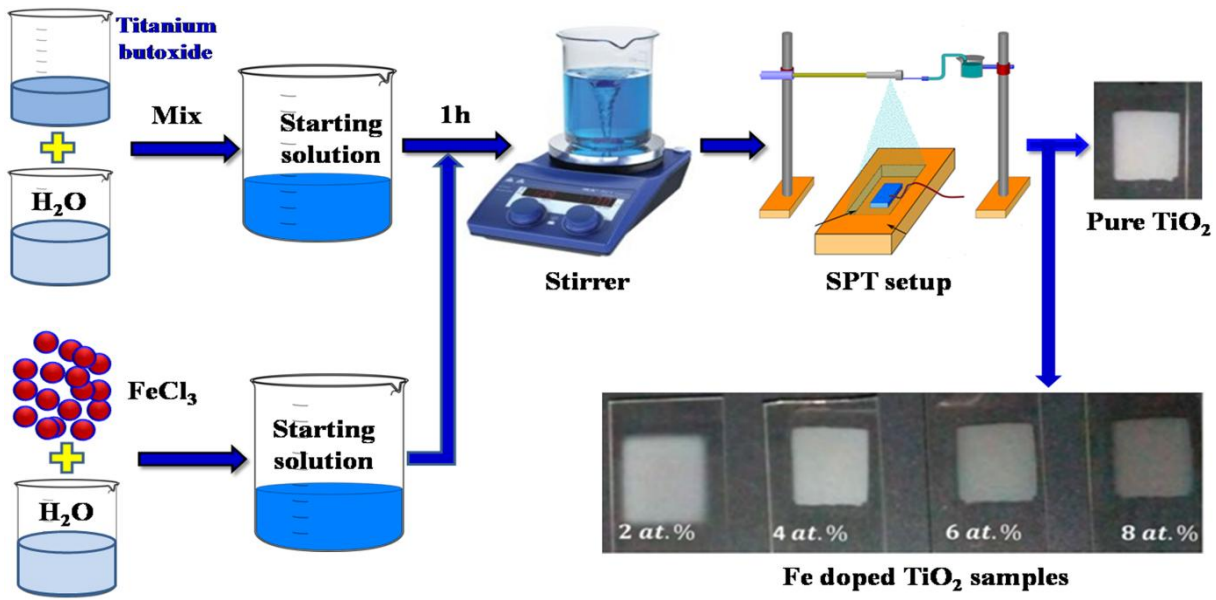

**Figure 2. A plausible diagram of pristine and Fe-doped TiO$_2$ thin films**

The formation of metal oxide (TiO$_2$) particles has been occurred by hydrolysis of metal alkoxides. The following overall reaction between Ti (IV)-butoxide (precursor) and water (reactant) that could occur on hot substrate resulting in the formation of TiO$_2$ thin film are given as follows.

$$Ti(OC_4H_9)_4 + 4H_2O \rightarrow Ti(OH)_4 + 4C_4H_9OH \quad (1)$$

TiO$_2$ is, then, formed by the condensation of the hydrolyzed species.

$$Ti(OH)_4 \rightarrow TiO_2 + 2H_2O \uparrow \quad (2)$$

The possible chemical reactions during the formation of Fe doped TiO$_2$ thin films are given below.

$$FeCl_3 + 3H_2O \rightarrow Fe^{3+} + 3HCl + 3(OH)^- \quad (3)$$
$$Ti(OH)_4 + Fe^{3+} \rightarrow TiO_2{:}Fe + H_2O \uparrow \quad (4)$$

Hence, the formation of corresponding oxide associated with Fe might not occur in the present research.

## 2.3 Characterization techniques

The resultant pristine and Fe doped thin films are undergone for their structural and optical characterizations. XRD diffractometer (3040XPertPRO, Philips) has been used to examine the crystallinity and phase identification. The absorbance spectra of the products are measured using a UV–Visible–NIR spectrophotometer.

## 3 Results and discussion

### 3.1 Microstructural investigations

**Figure 3** displays the XRD patterns of all thin film samples of $TiO_2$, doped with Fe content of 0.0, 2.0, 4.0, 6.0 and 8.0 at.%. The obtained XRD peaks in **Fig. 3a** confirm the anatase phase of tetragonal crystal arrangement in pristine sample having space group I41/amd(JCPDS 21-1272) [28] and highly intensed peak indicates the good crystallinity of $TiO_2$ nanoparticles. The addition of 2 at.% Fe does not alter the lattice structure and phase. After that, with the introduction of further impurity level (Fe content i.e., 4, 6 and 8 at.% Fe) the pristine product is noticed to exhibit the mixed phase of $TiO_2$ and transform from anatase to rutile phase (**Figure 3a**).

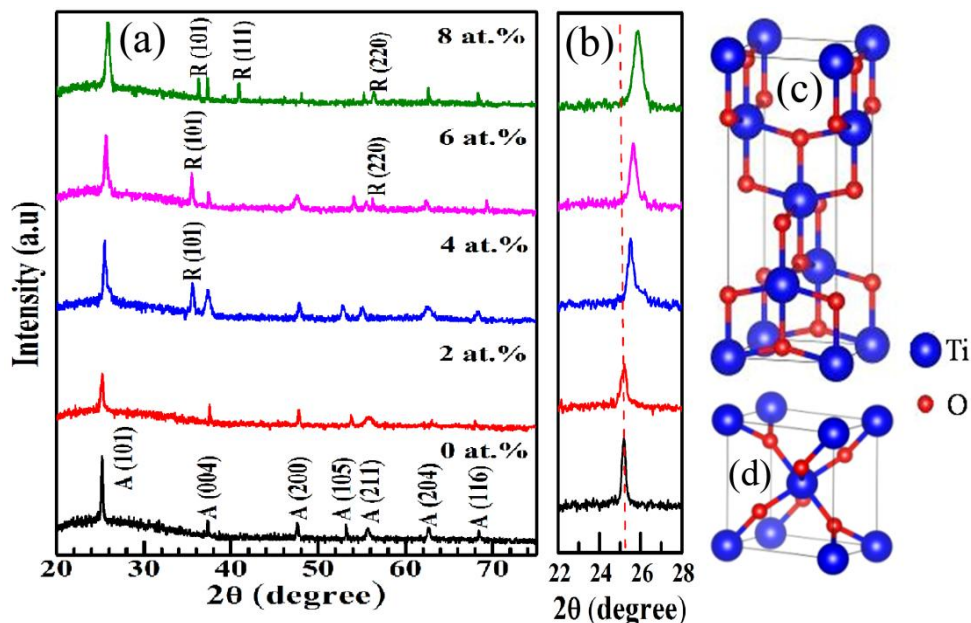

**Figure 3.** (a) XRD patterns of pristine and Fe doped $TiO_2$ thin films. (b) Magnification view of (101) plane (c) anatase and (d) rutile phases of $TiO_2$

The $TiO_2$ nanoparticles has constructive growth along (1 0 1) plane appearing preferential orientation along (1 0 1). The dominating peak is observed to shift toward higher 2θ positions as depicted in Fig. 3b confirming the conversion of anatase to rutile phase due to Fe content [29]. The anatase and rutile crystal structure of TiO2 has been depicted in Fig. 3c and 3d, respectively.

### 3.1.1 Calculation of D using Scherrer formula:

To study the nanocrystalline properties of the material it is very crucial to study the crystallite size (D) and lattice strain ($\varepsilon$). Numerous techniques have been developed to estimate these characteristics. For the estimation of D values of bare as well as Fe-doped $TiO_2$, firstly we used FWHM, and the peak of the planes from the XRD data in the Debey-Scherrer equation [30],

$$D_S = \frac{K\lambda}{\beta_D \cos\theta} \quad (5)$$

Where $D_S$ = crystallite size, $\lambda$ = wavelength of x-ray radiation ($\lambda$=1.54056 Å for Cu-K$\alpha$), $\beta_D$ = half-width full maximum, $\theta$ = Bragg's angle in degree and K=0.09. The estimated values of $D_S$ are shown in Table 2. The variation of the crystallite size D and dislocation density has shown in Fig 4. It maintains the inverse relation. As the Fe contents increase in $TiO_2$ lattice the crystallite size decreases and dislocation density displays the opposite fashion.

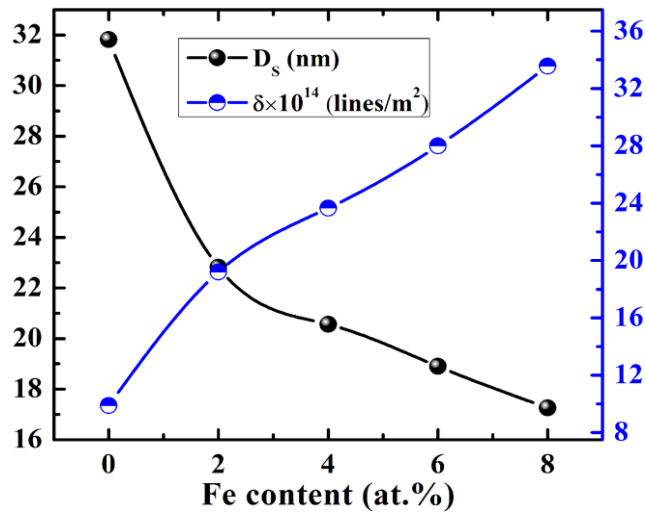

**Figure 4. Variation of D obtained from Scherrer method and dislocation density as a function of Fe content**

### 3.1.2 Measurement of D and $\varepsilon$ using W-H technique

We have applied W-H technique to evaluate the value of the crystallite size and strain is the Williamson and Hall (W-H) method written as [31]

$$\beta\cos\theta = \frac{K\lambda}{D_{WH}} + 4\varepsilon_{WH}\sin\theta \quad (6)$$

Where $\varepsilon_{WH}$ is the strain and $D_{WH}$ is the crystallite size. The above relation (6) is a linear equation like y= mx+c. We have taken $\beta_{hkl}\cos\theta$ in the y-axis and $4\sin\theta$ is taken in the x-axis to fit the W-H plot. We have evaluated strain from the slope and crystallite size from the y-intercept. The relationship of $D_{WH}$, $\varepsilon_{WH}$ and $\delta$ with Fe as doping element has been depicted in Fig 5.

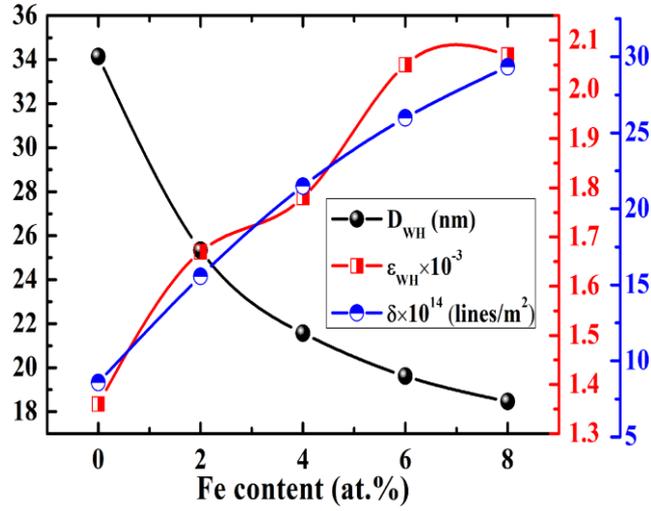

**Figure 5.** Variation of crystallite size, strain and dislocation density obtained from W-H plot.

### 3.1.3 Measurement of D and ε using H-W technique

The third method we have adopted for the estimation of D and ε is the Halder-Wagner (H-W) method that is expressed as [15].

$$\left(\frac{\beta}{tan\theta}\right)^2 = \frac{K\lambda}{D_{HW}} \cdot \frac{\beta}{tan\theta sin\theta} + 16\varepsilon_{HW}^2 \qquad (7)$$

Where $D_{HW}$ and $\varepsilon_{HW}$ define the size and strain, respectively. The above equation is a linear plot of $\left(\frac{\beta}{tan\theta}\right)^2$ vs $\frac{\beta}{tan\theta sin\theta}$ that is known as the H-W plot. We have determined $D_{HW}$ from the slope of the straight-line plot and $\varepsilon_{HW}$ from the y-intercept of the plot. The variation of $D_{HW}$, $\varepsilon_{HW}$ and $\delta$ with Fe contents as a dopant has been delineated in Fig.6.

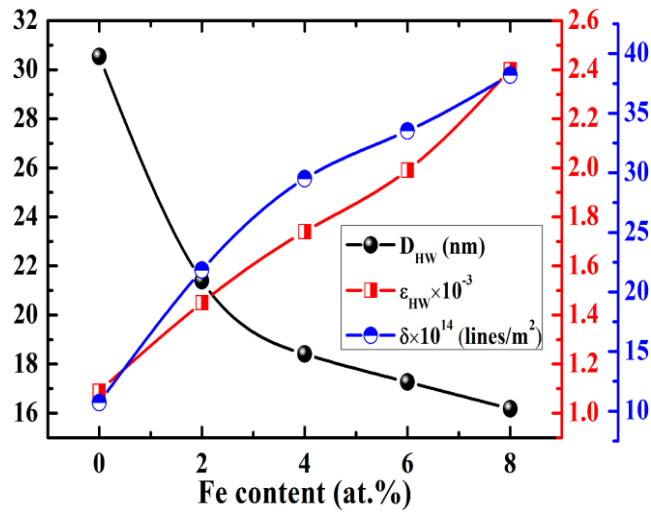

**Figure 6.** Variation of crystallite size, strain and dislocation density obtained from H-W plot.

### 3.1.4 Numerous microstructural parameters

The strain (ε) can be defined from the Stokes–Wilson formula[32]

$$\varepsilon = \frac{\beta_D}{4 \tan \theta} \quad (8)$$

in the above equation $\beta_D$ =half with full maximum, and θ= Bragg's angle.

The dislocation density (δ) was determined by using the following equation[33]

$$\delta = \frac{1}{D^2} \quad (9)$$

here D = crystallite size.

Microstress($\sigma_s$) has been computed using the following relation [34]

$$\sigma_s = \frac{\varepsilon}{2} E \quad (10)$$

In this equation ε= strain and E=282.76 GPa, Young modulus of the titanium oxide.

The interplanar spacing (d) can be estimated from Bragg's relation.

$$2d\sin\theta = n\lambda \quad (11)$$

The lattice parameter a=b and c can be estimated by the formula,

$$\frac{1}{d^2} = \frac{h^2 + k^2}{a^2} + \frac{l^2}{c^2} \quad (12)$$

(h, k, l)= Miller indices for Bragg's plane.

**Table 2. Deviation of crystallite size and strain of the resulted TiO$_2$ products determined applying different methods.**

| Fe contents (at.%) | Microstress (GPa) | Strain $\varepsilon \times 10^{-3}$ | Scherrer formula $D_S$ (nm) | W-H technique | | H-W technique | |
|---|---|---|---|---|---|---|---|
| | | | | $D_{WH}$ (nm) | $\varepsilon_{WH} \times 10^{-3}$ | $D_{HW}$ (nm) | $\varepsilon_{HW} \times 10^{-3}$ |
| 0 | -0.7041 | 4.98 | 31.82 | 34.14 | 1.36 | 30.54 | 1.09 |
| 2 | -0.7903 | 5.59 | 22.82 | 25.34 | 1.67 | 21.40 | 1.45 |
| 4 | -0.8214 | 5.81 | 20.56 | 21.57 | 1.78 | 18.41 | 1.74 |
| 6 | -0.8865 | 6.27 | 18.90 | 19.62 | 2.05 | 17.27 | 1.99 |
| 8 | -1.6895 | 11.95 | 17.26 | 18.46 | 2.07 | 16.18 | 2.40 |

The texture coefficient (TC) is estimated employing the equation below [35]

$$TC_{(hkl)} = \frac{I_{(hkl)}/I_{0(hkl)}}{\frac{1}{n}\sum I_{(hkl)}/I_{0(hkl)}} \quad (13)$$

Where $I_{(hkl)}$ = relative intensity and $I_{0(hkl)}$ = standard intensity of the (hkl) plane.

The values of stacking fault (SF) can be calculated using the following relation

$$SF = \left[\frac{2\pi^2}{45(3\tan\theta)^{1/2}}\right]\beta \quad (14)$$

The estimated values of the structural parameters are shown in Table 2 and Table 3

**Table 3. Deviation of mumerous structural parameters of bare and Fe-incorporated products**

| Fe contents (at.%) | $\delta$ (lines/m$^2$)×10$^{14}$ | | | Lattice constants (Å) | | Texture Coefficients | Stacking Fault |
|---|---|---|---|---|---|---|---|
| | Scherrer formula | W-H technique | H-W technique | a = b | c | | |
| 0 | 9.88 | 8.58 | 10.72 | 3.81 | 9.43 | 1.0829 | 0.002392 |
| 2 | 19.20 | 15.57 | 21.83 | 3.79 | 9.67 | 1.3334 | 0.003334 |
| 4 | 23.66 | 21.49 | 29.50 | 3.75 | 9.51 | 1.4318 | 0.003680 |
| 6 | 28.00 | 25.98 | 33.52 | 3.74 | 9.36 | 1.5187 | 0.003995 |
| 8 | 33.57 | 29.35 | 38.20 | 3.69 | 9.59 | 1.4548 | 0.004357 |

### 3.2 Dispersion of optical constants

**Figure 7** presents the UV–vis absorption spectra of TiO$_2$ and Fe doped TiO$_2$ samples as a function of $\lambda$ in the range of 300-1200 nm and exhibits the sharp absorption edges in the UV light regionat around 450nm.The magnified result of absorbance upto 0.3 is depicted in the **Fig. 7**. The deposited TiO$_2$ products show an average absorption (0.12) between 500 nm and 1200 nm. The absorbance of 2 at.% Fe doped TiO$_2$ illustrates an enhancement characteristic and declines for the higher doping levels (4, 6 and 8 at.%). This reduction of absorbance values of doped products might be attributed to the increase of photon scattering by lattice defects induced from Fe impurity and result of the despair and passing of electrons from the VB to CB of TiO$_2$ crystal [36]. The Fe doping induces a red shift in the absorption edges i.e., the absorption edge is shifted towards the lower photon energy, which indicates the decrement in band gap. The shift of absorption edges appears in the iron doped TiO$_2$ products may be owing to the transfer of electrons from valence band to the conduction band of TiO$_2$ [37].

In order to evaluate the bandgap ($E_g$) values using the absorbance data, the plots of $(\alpha h\nu)^2$ as a function of $h\nu$ have been drawn in **Fig. 8** and expressed by Tauc's relationship [23]:

$$\alpha h\nu = A(h\nu - E_g)^{1/2} \quad (15)$$

where α is the absorption coefficient, A impliesa constant, n implies an exponent, n = 1/2 for direct transition and n = 2 for indirect transition.

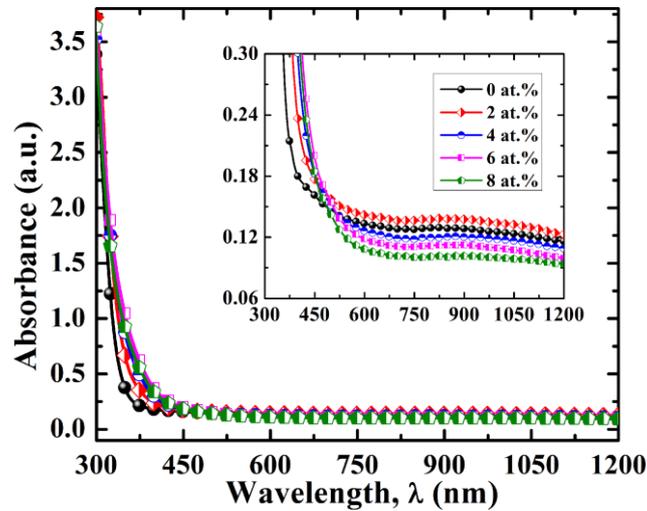

**Figure 7. Wavelength dependent absorption behavior of pristine and Fe-incorporated titanium dioxide samples**

We have given an attention to find the direct bandgap because of the values of n remains fixed inside the intended energy range [38]. The thin film's bandgap values have been determined as a result of extrapolating the linear part of the plotstoward lower energies. The intercepts on hν axis provides the $E_g$ values which has varied from 3.70 eV to 3.81eV. The obtained results of $E_g$ in the current study are comparable to the range of $E_g$ of $TiO_2$ reported in the literature [39]. The $E_g$ values are influenced by the crystallinity of products. When the crystallitesize decreases, the density of the grain boundaries increases and,as a result, more carriers are trapped in the space chargeregion, leading to a decrease in the free carriers. In addition, this decrement in $E_g$ may be associated to the trapping level formed between the VB and CB of $TiO_2$ by Fe impurity (**Fig. 9**) [1][37].

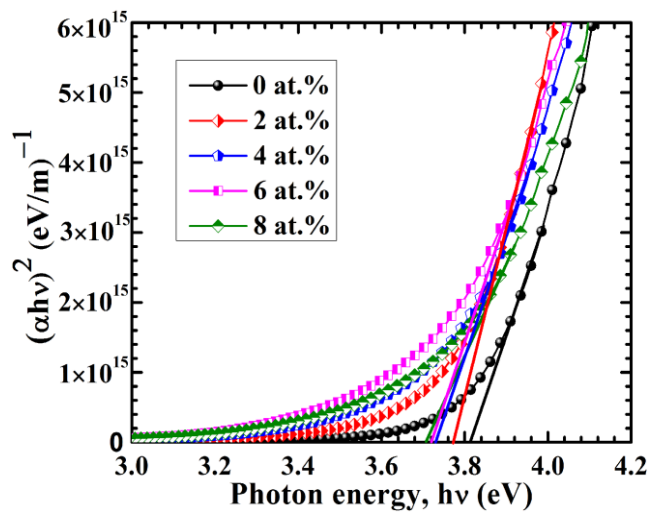

**Figure 8. $E_g$ of pristine and Fe doped $TiO_2$ thin films.**

Refractive index (n) of the resultant pristine and Fe-doped products is obtained using the following relation [40].

$$n = \left(\frac{1+R}{1-R}\right) + \sqrt{\left(\frac{4R}{(1-R)^2} - k^2\right)} \qquad (16)$$

In this above euation (16), R is the reflectance of the $TiO_2$ thin film which is calculated from the T and A data.

The extinction coefficient (k) is also known as the attenuation coefficient, the loss of light energy per unit penetration in the medium, which also describes that amount of the light energy is absorbed when light travels through that medium. For the absorbing medium, the value of k is much higher than the transparent medium [36].

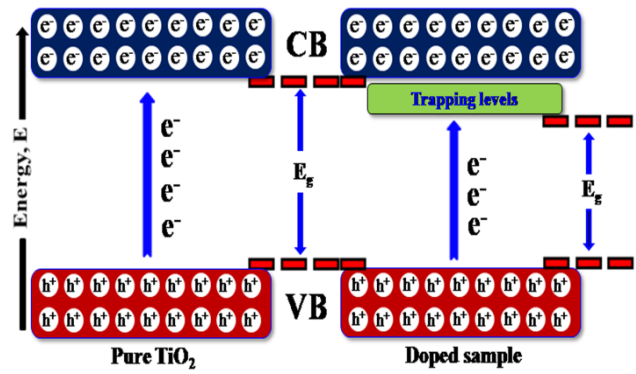

**Figure 9. A hypothetical analysis of band gap structures of pure and Fe doped $TiO_2$**

The extinction coefficient (k) of the prepared pure and doped samples can be measured using the following equation [41].

$$k = \frac{\alpha \lambda}{4\pi} \qquad (17)$$

where α is the absorption coefficient, and λ is the wavelength of the radiation.

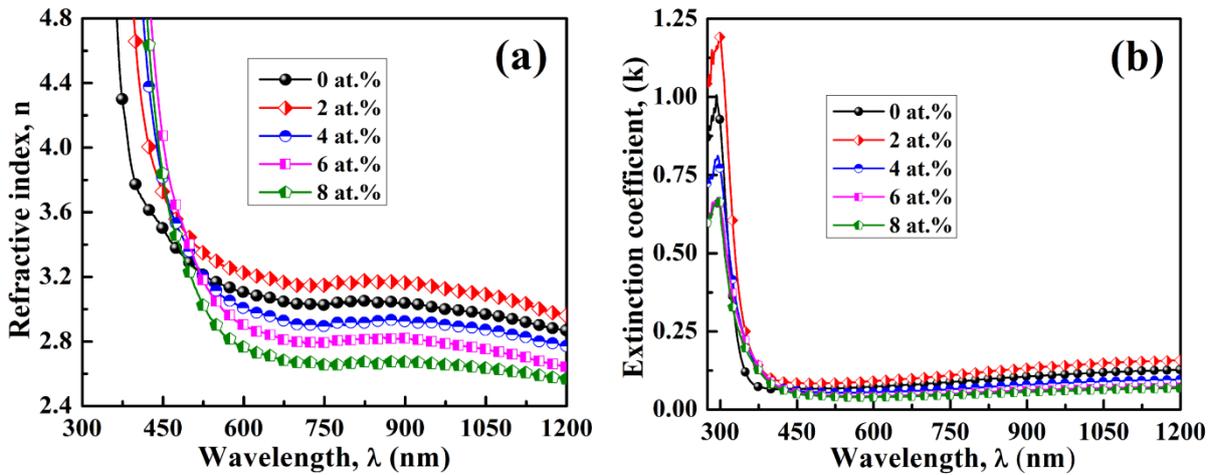

**Figure 10. (a) Deviation of n and (b) k with λ for Fe doped $TiO_2$ thin films with various Fe concentrations**

The variation of refractive index (n) and extinction coefficient (k) with wavelength λ is given in Figure 10. In the case of 2% Fe-doped TiO$_2$ film the value of n and k increases. This can be explained as the decrease of the grain size when Fe is incorporated [42]. After that with the increase of Fe content in the sample (4, 6, 8 at.% Fe-doped TiO$_2$), n and k are noticed to decline. This decrease of n and k can be subjected to the increase of carrier concentration in the highly Fe-doped thin films [43]. In addition, the value of k decreases with the increase of Fe doping owing to the improvement of surface smoothness which implies the increment of the transmitting capability (as, A+T+R=1). Hence absorbance decreases as Fe (4%, 6%, 8%) content increases in the TiO$_2$ thin film [44].

The complex dielectric constant (ε) expresses the relation between electromagnetic rays with matter [45]. ε equals to ($\varepsilon_r + i\varepsilon_i$), where $\varepsilon_r$ is the real part and $\varepsilon_i$ is the imaginary part of dielectric constant expressed through the given relations,

$$\varepsilon_r = n^2 - k^2 \qquad (18)$$

$$\varepsilon_i = 2nk \qquad (19)$$

A plot of $\varepsilon_r$ and $\varepsilon_i$ with λ is shown in fig. 11. $\varepsilon_r$ is related to the speed of light in the medium and $\varepsilon_r$ is related to the transparency of medium. Both $\varepsilon_r$ and $\varepsilon_i$ have higher values in 2% Fe doped samples. This can be ascribed as the light absorption increases because of dipole motion or free carriers absorption [46]. From the diagram 11, it is clearly seen that $\varepsilon_r \gg \varepsilon_i$. The overall lower value of ε means that the material is more transmissive [43][47][48]. The nature of $\varepsilon_i$ has a similar trend as of k which is related to the absorbance [49].

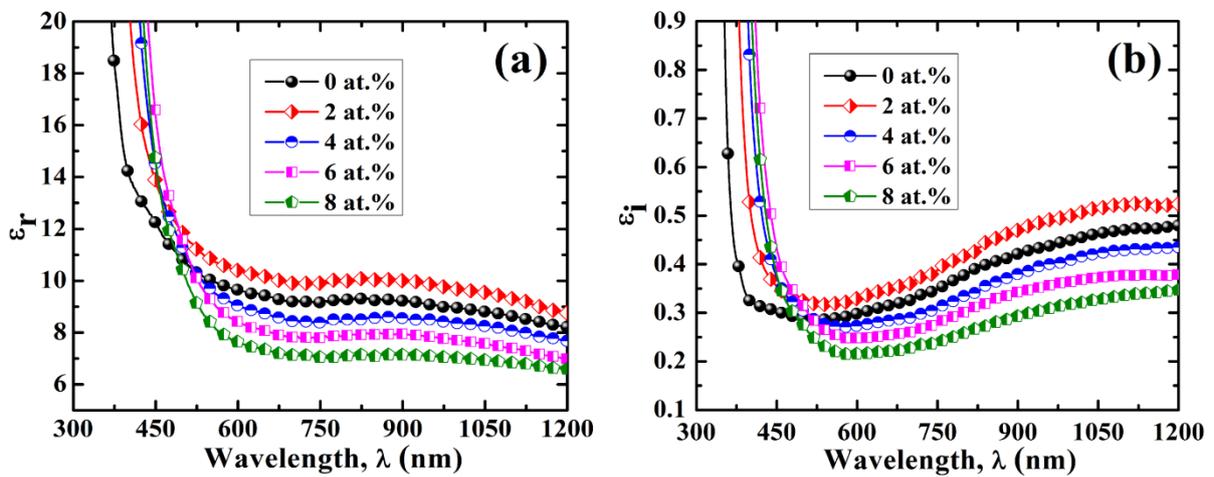

Figure 11. Plot of (a) $\varepsilon_r$ and (b) $\varepsilon_i$ of ε with wavelengths

The knowledge and data about the real and imaginary parts of the complex dielectric constant give the value of loss factor (tanδ) which is defined as.

$$\tan\delta = \frac{\varepsilon_i}{\varepsilon_r} \qquad (20)$$

tanδ is displayed with wavelength in Fig. 12 which reveals that the loss factor increases as λ increases, i.e., when light with high energy incident on the material the loss factor decrease [12][50]. At low frequency, the dielectric loss is very high because of the presence of electronic, ionic, dipolar, and space charge polarization [51]. The loss factor increases in the case of 2% Fe adulteration in $TiO_2$ due to the random distribution of Fe contents in $TiO_2$ crystal framework [46].

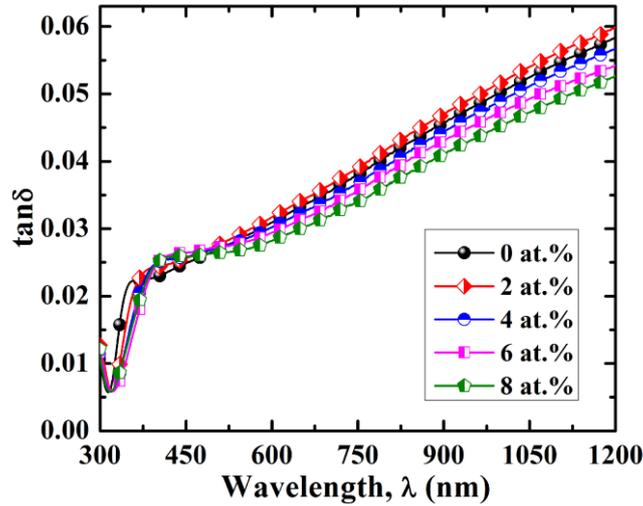

**Figure 12. Deviation of tanδ as a function of λ**

The decrease of loss energy with the increase of Fe-content (4%-8%) in the samples which implies that carriers' concentration increases and they absorb more energy [42].

The real and imaginary dielectric constants are also used to determine the characteristics of the volume energy loss factor (VELF) and surface energy loss factor (SELF). The relations are given below,

$$\text{VELF} = \frac{\varepsilon_i}{\varepsilon_r^2 + \varepsilon_i^2} \quad (21)$$

$$\text{SELF} = \frac{\varepsilon_i}{(\varepsilon_r + 1)^2 + \varepsilon_i^2} \quad (22)$$

The plots of VELF and SELF are shown in the Fig. 13. VELF and SELF inform us about the probability of energy loss of fast-moving electrons through the bulk and surface of the material [52][53]. From Fig.13, it is observed that VELF and SELF have a similar pattern and their values decrease as photon energy increases. VELF is greater than SELF for both pristine and Fe-doped $TiO_2$. This occurs because the motion of electrons within the samples relies on the amalgamate of cluster [54].

The relation among refractive index (n), lattice dielectric constant ($\varepsilon_L$) and wavelength (λ) can be written as, [47]

$$\varepsilon = n^2 = \varepsilon_L - \left(\frac{e^2 N}{4\pi c^2 \varepsilon_0 m^*}\right)\lambda^2 \quad (23)$$

Where c = the velocity of light, $\varepsilon_0$ = DC in the free space ($\varepsilon_0 = 8.854 \times 10^{-12}$ F/m), e = electric charge. In formula (23) $N/m^*$ can be expressed as

$$\frac{N}{m^*} = \frac{free\,carrier\,density}{free\,carrier\,effective\,mass} \qquad (24)$$

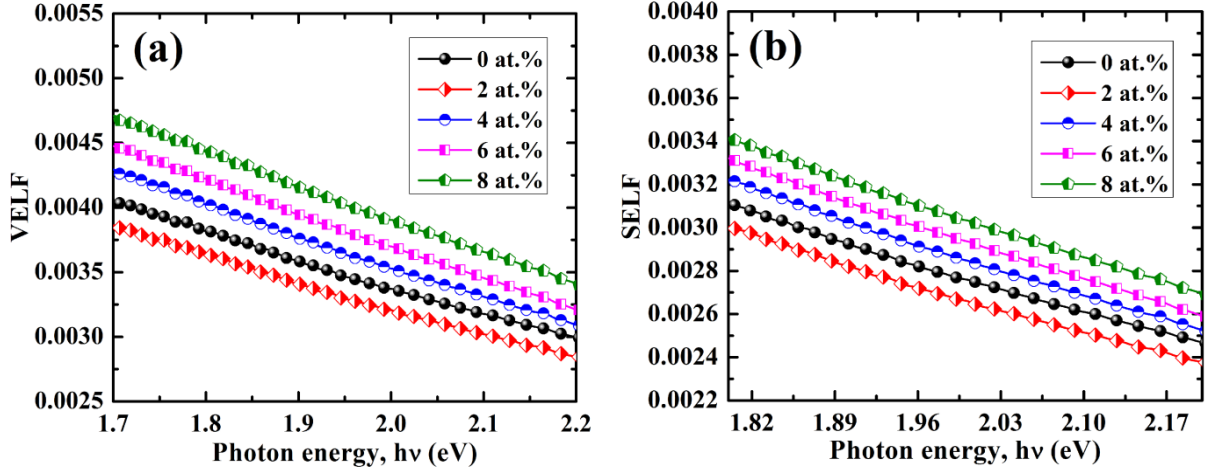

**Figure 13. Variation of (a) VELF and (b) SELF with hν**

Fig. 14 reveals the linear relationship of $n^2$ with $\lambda^2$ for different Fe contents. The y-intercept (y=0) gives the value of $\varepsilon_L$ which is noticed to increase after 2 at.% Fe impurity and then decreases as the Fe content increases (Table 4).

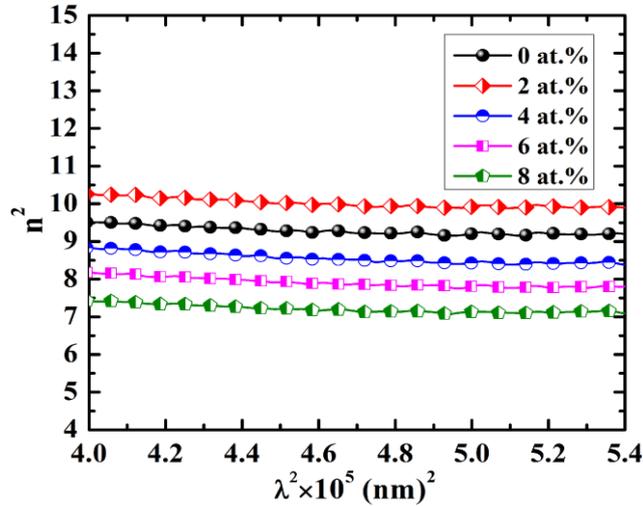

**Figure 14. Relation between $n^2$ with $\lambda^2$**

To get the idea of density and structure of the material it is essential to find the values of static refractive index ($n_0$) [47]. The high-frequency dielectric constant, $\varepsilon_\infty$ can be evaluated from the static refractive index, $n_0$ which is known as the Sellmeier oscillator [52].

$$\frac{n_0^2 - 1}{n^2 - 1} = 1 - \left(\frac{\lambda_0}{\lambda}\right)^2 \qquad (24)$$

Where $\lambda_0$ is the average oscillator wavelength. Figure 15 depicts the plot of $(n^2-1)^{-1}$ vs $\lambda^{-2}$ which reveals a linear equation like y= mx+c, where $\lambda^{-2}$ and $(n^2-1)^{-1}$ have been drawn in the horizontal and vertical directions, respectively. From the y-intercept, we can determine the static refractive index, $n_0$. Moreover, by evaluating the slope of the line we can calculate the average oscillator wavelength $\lambda_0$. The high frequency dielectric constant, $\varepsilon_\infty$ can be estimated by using the following relation [12]

$$n_0 = \sqrt{\varepsilon_\infty} \qquad (25)$$

In addition, the average oscillator energy ($S_0$) has been evaluated from the following relation [45]

$$S_0 = \frac{n_\infty^2 - 1}{\lambda_0^2} \qquad (26)$$

The estimated values of $S_0$, $\lambda_0$, $n_0$, $\varepsilon_L$, and $\varepsilon_\infty$ are tabulated in the Table 4 and their variations are depicted in the Fig. 16. It has been noticed from the Fig. 16 all the parameters follows the similar trends [12][47]. From the Table 4 it is observed that the value of $S_0$ is maximum for 2% Fe doped samples. It can be attributed that high value of $S_0$ means high crystalline behaviour among the Fe doped samples [47].

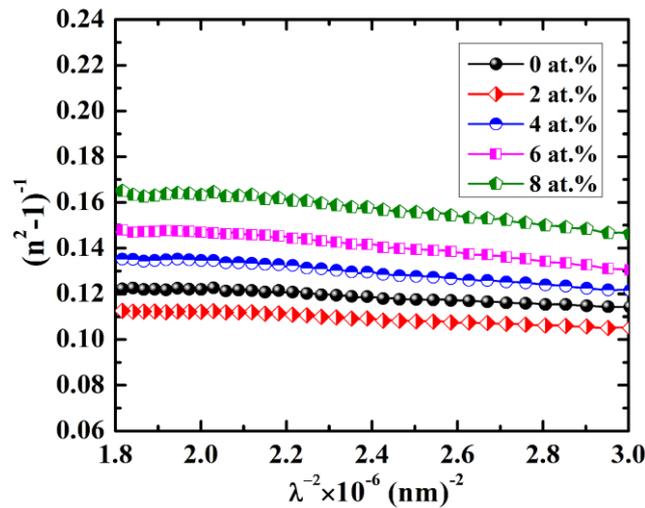

**Figure 15. Relation between $(n^2-1)^{-1}$ and $\lambda^{-2}$ of TiO$_2$ films**

**Table 4. Numerous optical dielectric parameters of pristine and Fe-doped titanium dioxide.**

| Fe contents (at.%) | $\varepsilon_L$ | $\varepsilon_\infty$ | $n_0$ | $\lambda_0$ (nm) | $S_0 \times 10^{13}$ | $N/m^* \times 10^{20}$ (m$^{-3}$kg$^{-1}$) |
|---|---|---|---|---|---|---|
| 0 | 10.38 | 8.21 | 2.87 | 244.71 | 12.09 | 1.46 |
| 2 | 11.21 | 8.85 | 2.97 | 244.32 | 13.10 | 1.60 |
| 4 | 10.01 | 7.19 | 2.68 | 288.70 | 7.42 | 1.94 |
| 6 | 9.21 | 6.47 | 2.54 | 307.90 | 5.75 | 1.73 |
| 8 | 8.24 | 5.95 | 2.44 | 302.32 | 5.42 | 1.37 |

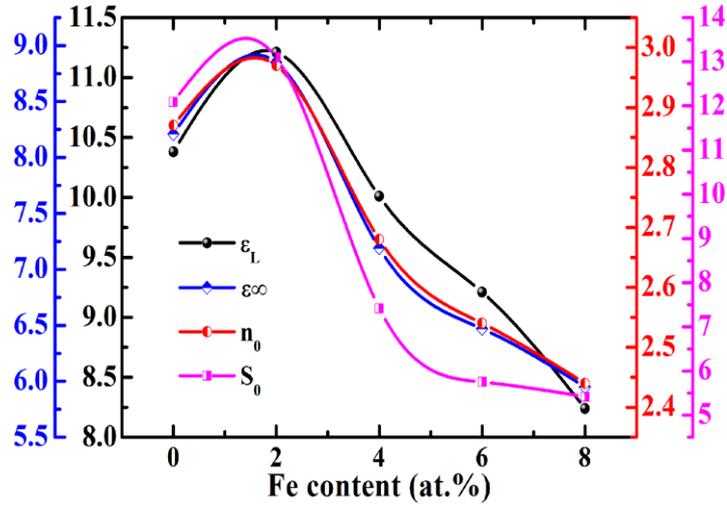

**Figure 16.** A comparison of $\varepsilon_L$, $\varepsilon_\infty$, $n_0$ and $S_0$ as a function of Fe contents.

### 3.3 Dispersion behavior of TiO$_2$ thin films:

It is very crucial to study the dispersion parameters of TiO$_2$ thin films as it gives important ideas to design the practical devices of optical communication and spectral dispersion. It is obvious to know the dispersive behabiour of TiO$_2$ to understand whether TiO$_2$ thin films are potential candidate for practical uses. We have employed the Wemple and DiDomenico (W–D) model. According to this model refractive index is plotted according to the following relation [44][55]:

$$n^2 - 1 = \frac{E_0 E_d}{E_0^2 - (h\nu)^2} \qquad (27)$$

Where $E_0$ is the single oscillator energy and $E_d$ is the dispersion energy parameter which measures the average bandgap energy ($E_0 = 1.5 E_g$) and the average strength of the interband optical transitions respectively. The energy dispersion parameter $E_d$ is connected with the harmonization of positive ions closed to each other, ionicity, anion valency, and the efficient quantity of diffusion electrons [47][56]. Considering the above equation (27), a diagram of $(n^2 - 1)^{-1}$ vs $(h\nu)^2$ has been depicted in Fig. 17. By estimating the y-intercept and slope from the $(n^2 - 1)^{-1}$ vs $(h\nu)^2$ plot the value of $E_d$ and $E_0$ are calculated, respectively.

The highest value of $E_d$ has been noticed for 2 at.% Fe doped sample which is related to the structural changes of the thin film [47]. The value of $E_d$ decreases (for 4, 6 and 8 at.% of Fe content) which implies that the co-ordination number of atoms decreases. The value of $E_0$ decreases with the increase of Fe element as a dopant in TiO$_2$ which may be due to the increase of scattering centre and the decrease of $E_g$ [12].

In oredre to get a better understanding and knowledge of the ionic bond strength of a material it is very important to investigate another energy parameter known as lattice energy ($E_1$). Lattice energy $E_l$ related to the above discussed energy parameters $E_0$ and $E_d$ can be estimated from the equation [57].

$$n^2 - 1 = \frac{E_0 E_d}{E_0^2 - E^2} - \frac{E_1^2}{E^2} \qquad (28)$$

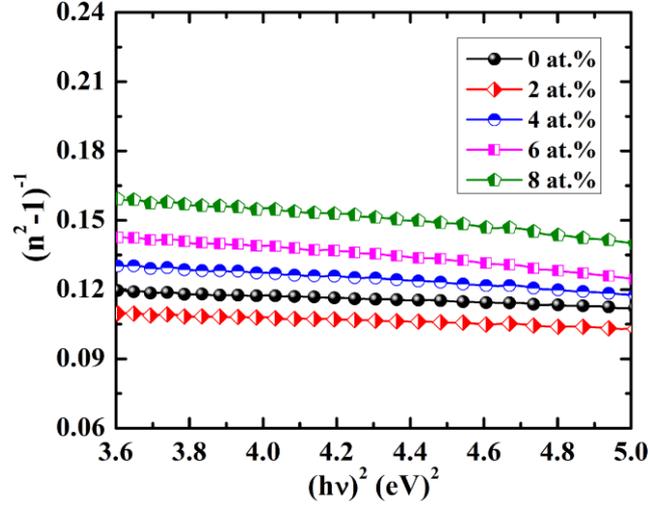

**Figure 17. Plot of $(n^2-1)^{-1}$ vs $(h\nu)^2$ for pristine and Fe-doped $TiO_2$**

In the lower energy region where the wavelength is longer than phonon resonance, means $E_0^2 \gg E^2$. $(n^2 - 1)$ is plotted against $(h\nu)^{-2}$ (Fig. 18) which will be a straight line according to Poignant [45][58] in the long-wavelength region. Considering this approximation one can write the above expression in the following form,

$$n^2 - 1 = \frac{E_d}{E_0} - \frac{E_1^2}{E^2} \qquad (29)$$

The y-intercept gives $\frac{E_d}{E_0}$ and the slope of the line is equal to $-E_1^2$. From the calculation of slope, the lattice energy parameter can easily be determined.

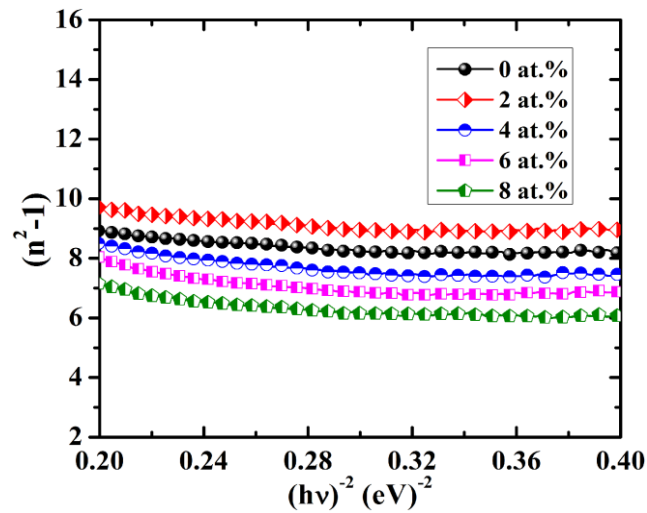

**Figure 18. Variation of $(n^2-1)$ with $(h\nu)^{-2}$**

**Table 5. Dispersion energy parameters of pristine and Fe-doped titanium dioxide samples**

| Fe contents (at.%) | $E_g$ (eV) | $E_d$ (eV) | $E_0$ (eV) | $E_1$ (eV) | $E_0/E_g$ |
|---|---|---|---|---|---|
| 0 | 3.81 | 38.25 | 5.25 | 2.31 | 1.38 |
| 2 | 3.76 | 41.86 | 5.27 | 2.40 | 1.40 |
| 4 | 3.72 | 26.51 | 4.30 | 2.74 | 1.16 |
| 6 | 3.71 | 20.71 | 3.90 | 2.83 | 1.05 |
| 8 | 3.70 | 19.26 | 3.98 | 2.59 | 1.08 |

### 3.4 Conclusion

In summary, a general overview of synthesis and chracterizations of pure and Fe doped $TiO_2$ thin films has been presented in this paper and cost effective spray pyrolysis route has been employed to synthesize the samples meticulously. XRD study shows the pure and Fe doped samples are nano structured. A phase transition, anatase phase to mixed phase has been confirmed from the XRD, when $TiO_2$ is adulterated by Fe (4%, 6% and 8%). A through investigation of microstructural parameters like crystallite size, strain, dislocation density are performed by D-S, W-H, and H-W methods. The investigated results reveals that when the doping concentration is increased from 0 to 8 at. % the crystallite size declined from $D_S$ = 31.82 to 17.26 nm, $D_{WH}$ = 34.14 to 18.46 nm and $D_{HW}$ = 30.54 to 16.18 nm, while the strain was found to decrease from $\varepsilon_{WH}$ = 1.36 to 2.07, $\times 10^{-3}$ and $\varepsilon_{HW}$ = 1.09 to 2.40, $\times 10^{-3}$, respectively. From the UV vis spectra, it is observed that the absorption edge of Fe-doped $TiO_2$ products is shifted towards longer wavelengths (i.e. red shifted) which causes the reduction of bandgap from 3.81 eV to 3.70 eV. Other optical parameters like dielectric constants, refractive index, energy loss factor, SELF, VELF and lattice energy are estimated. The Fe-doped $TiO_2$ thin films may be considered as a promising and effective candidate in designing and manufacturing electronic devices.

### Acknowledgments

We acknowledge the Department of Physics, Department of Glass and Ceramic Engineering, Bangladesh University of Engineering and Technology (BUET), Dhaka, Bangladesh, and also Bangladesh Atomic Energy Center, Dhaka, Bangladesh Atomic Energy Commission (BAEC) for their cordial supports.

### Declaration of interests

We certify that this manuscript has not yet been submitted to any journal and there are no affiliations with or involvement in any organization or entity with any financial interest in the subject matter or materials discussed in this manuscript.